# Absolute spectroscopy near 7.8 μm with a comb-locked extended-cavity quantum-cascade-laser

MARCO LAMPERTI[1,*], BIDOOR ALSAIF[2], DAVIDE GATTI[1], MARTIN FERMANN[3], PAOLO LAPORTA[1], AAMIR FAROOQ[2], AND MARCO MARANGONI[1]

[1]*Physics Department of Politecnico di Milano and IFN-CNR, Via G. Previati 1/C, 23900 Lecco, Italy*
[2]*King Abdullah University of Science and Technology (KAUST), Clean Combustion Research Center (CCRC), Thuwal 23955, Saudi Arabia*
[3]*IMRA America Inc., 1044 Woodridge Avenue, Ann Arbor, MI 48105-9774, USA*
*\*Corresponding author: marco1.lamperti@polimi.it*

**We report the first experimental demonstration of frequency-locking of an extended-cavity quantum-cascade-laser (EC-QCL) to a near-infrared frequency comb. The locking scheme is applied to carry out absolute spectroscopy of $N_2O$ lines near 7.87 μm with an accuracy of ~60 kHz. Thanks to a single mode operation over more than 100 $cm^{-1}$, the comb-locked EC-QCL shows great potential for the accurate retrieval of line center frequencies in a spectral region that is currently outside the reach of broadly tunable cw sources, either based on difference frequency generation or optical parametric oscillation. The approach described here can be straightforwardly extended up to 12 μm, which is the current wavelength limit for commercial cw EC-QCLs.**

*OCIS codes (120.3930) Metrological instrumentation; (140.3425) Laser stabilization; (300.6340) Spectroscopy, infrared*

Since their invention, optical frequency combs have revitalized the field of precision molecular spectroscopy, making it possible to achieve accuracies at the kHz or even sub-kHz level on absorption line centers [1-4]. In order to bring such a comb revolution to the point of redefining spectroscopic databases such as HITRAN [5], which are still mostly based on a pre-comb spectroscopy era, it is crucial to develop spectrometers that join an accurate frequency axis to a wide spectral coverage of > 100 $cm^{-1}$, which is the typical extension of absorption bands. This is easier to be performed in the near-infrared, thanks to the availability of commercial frequency combs and of a variety of widely tunable diode-laser-based solutions. In this respect, demonstrations of accurate broad line surveys have been given for acetylene, ammonia and water in a sub-Doppler regime [6-10] and more recently for carbon monoxide in a Doppler broadening regime [11].

In the mid-infrared (mid-IR) region, the development of such spectrometer is more challenging. A first requirement is the comb-referencing of the mid-IR probe laser: this has been obtained by a variety of approaches, such as down-conversion of the frequency comb to the mid-IR through difference frequency generation (DFG) [12,13] or optical parametric oscillation (OPO) [14], up-conversion of the probe laser to the near-IR through sum-frequency or second-harmonic generation (SFG/SHG) [15-18], as well as referencing schemes applied to DFG- and OPO-based cw sources [19,4]. A second requirement is a widely tunable laser source. Up to a wavelength of 4.5 μm, a viable solution is represented by cw sources based on DFG or OPO processes in periodically-poled lithium-niobate crystals: these have been exploited for sub-Doppler surveys over more than 50 $cm^{-1}$ on $CH_4$ lines near 3 μm and $N_2O$ lines near 4.5 μm [20]. Distributed-feedback QCLs are a valuable alternative, but only over a narrower spectral range, as demonstrated by Galli et al. [21] on $CO_2$ lines near 4.3 μm. The widest spectral coverage achieved so far was obtained by a dual-comb approach [22] that affords multi-parallel detection and extremely fast acquisition times: however, this comes at the price of an accuracy limited to ~300 kHz and of a setup composed of a pair of Hz-level-locked combs that can hardly be scaled for operation beyond 4.5 μm.

An extremely powerful alternative is represented by EC-QCLs: these enable single mode emission and frequency tuning in the mid-IR (from 4 to 12 μm) over ranges in excess of 100 $cm^{-1}$, with a 100 mW optical power. Their adoption for precision spectroscopy has been hampered so far by a large amount of frequency noise, resulting in an optical linewidth of ~ 15 MHz over 50 ms [23]. This is one of the reasons why neither their frequency nor their phase has been so far locked to a frequency comb. Their use in combination with frequency combs has been demonstrated by the group of N. Newbury in an open loop regime [24], which exploited the inherently fast and wide mode-hop-free tunability of these lasers, yet this approach could not reach an accuracy better than 800 kHz.

In this Letter, we report for the first time frequency locking of an EC-QCL to a near-IR frequency comb, the former at around 7.87 μm, the latter at 1.9 μm from a Tm:fiber oscillator. The locking is obtained by slow feedback to the EC-QCL piezo with a 100 Hz servo bandwidth, which results in a 100 kHz frequency stability over 100 ms. In these conditions, $N_2O$ absorption spectra can be acquired and fitted with an overall uncertainty of about 60 kHz on the line center frequency. The addition of a fast feedback loop acting on an external acousto-optic frequency shifter is also discussed: this allows a narrowing of the laser emission line by a factor of 8, but it introduces a severe laser intensity noise that makes this choice counterproductive for the spectrometer performance.

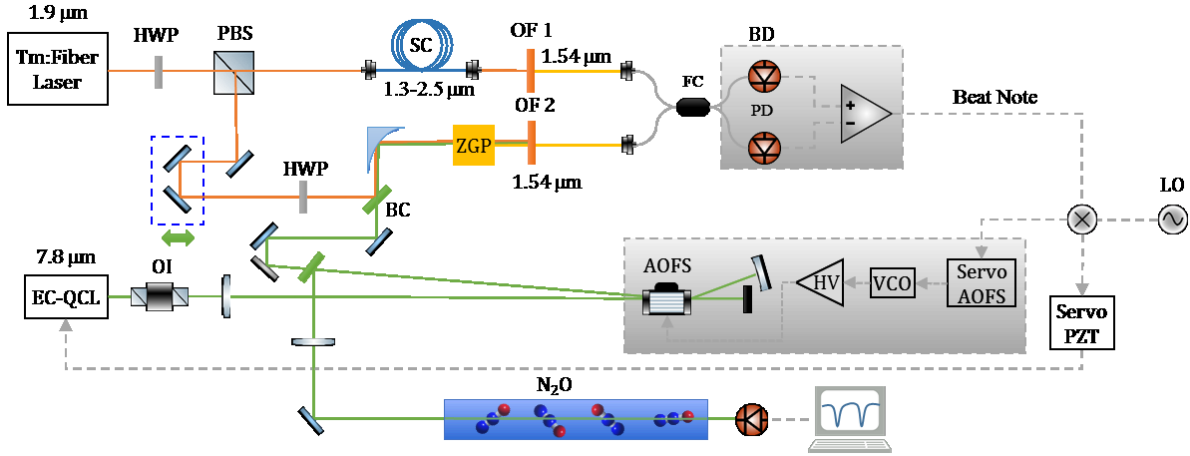

Fig. 1. Experimental setup. Green line: EC-QCL optical path. Orange line: Tm-comb optical path. Grey dashed lines: electrical links. OI: optical isolator. SC: supercontinuum. HWP: half-wave-plate. BC: beam combiner. OF: optical filter. FC: fiber coupler. BD: balanced detector. LO: local oscillator. AOFS: acousto-optic-frequency shifter. HV: high voltage amplifier. VCO: voltage-controlled oscillator.

The layout of the spectrometer is sketched in Fig. 1. The near-IR frequency comb is based on an amplified Tm-fiber oscillator at 100 MHz delivering up to 1.5 W at 1.9 µm. The EC-QCL (from Daylight Solutions) operates at room temperature and provides single-mode emission in the 7.55-8.2 µm range with an output power up to 50 mW after optical isolation. The referencing scheme relies on an SFG process [25], where part of the Tm output (100 mW) and of the EC-QCL (16 mW) are collinearly combined and focused by an off-axis parabolic mirror into an 8 mm long Zinc-Germanium Phosphide (ZGP) crystal. This gives rise to a new comb at around 1.54 µm, hereafter called SFG comb. The frequency of the SFG comb is offset from the original comb by the EC-QCL frequency. By heterodyning the SFG comb against a spectrally broadened replica of the near-IR comb, a radio-frequency (RF) beat note ($f_{beat}$) is eventually extracted, $f_{beat} = |f_{QCL} - m \cdot f_{rep}|$, which allows the EC-QCL frequency ($f_{QCL}$) to be determined against an integer multiple of the comb repetition frequency ($f_{rep}$). Differently from [24], where $f_{beat}$ was tracked in real time by fast digitization followed by fast-Fourier-Transform, $f_{beat}$ was here steadily locked to a local RF oscillator.

Two locking schemes have been implemented and tested. A first scheme makes use of one servo only, providing feedback to the EC-QCL through the available piezo modulation port. Due to a bandwidth limit of 100 Hz, it was not possible by this approach to go beyond a simple frequency locking. In a second scheme, to achieve a faster frequency correction and to explore the feasibility of phase locking, we added a second servo acting on an acousto-optic-frequency-shifter (AOFS). We recurred to external frequency actuation rather than to laser current control because for our laser the current tuning was high-passed at 10 kHz by the manufacturer, thus inhibiting any laser frequency control in the 100 Hz-10 kHz range. As sketched in the figure, the SFG branch of the setup is aligned to the beam diffracted by the AOFS. This was arranged in a double pass configuration to benefit from doubled frequency shifts while suppressing misalignments due to the changing diffraction angle.

Figure 2(a) reports the beat note spectrum in a free-running regime acquired with a sweep time of 6 ms at a 50 kHz resolution bandwidth. It can be noticed that the ECQCL frequency experiences a jitter of about 20 MHz at the ms timescale. The satisfactorily high signal-to-noise ratio (SNR) of more than 30 dB derives from an efficient nonlinear interaction in ZGP, which leads to an SGF comb power of ~80 nW, and from the use of balanced detection, which gives a reduced intensity noise floor and a 3 dB higher SNR as compared to direct detection. Figure 2(b) reports the averaged profiles of the beat note under slow (blue) and fast (red) locking. In the slow case, the linewidth suffers from the large frequency excursions of the laser beyond 100 Hz, thus beyond the available control bandwidth. This leads to a linewidth of 21 MHz, which is slightly worse than the 15 MHz value reported in [23] for a 4.5 µm EC-QCL. The addition of a second faster feedback loop better compensates for the laser frequency jitter and narrows the emission line down to 2.5 MHz (Fig. 2(b)), which corresponds to an improvement by a factor of 8 in the available spectral resolution once the laser is applied to spectroscopy. The electrical spectra of the error signal reported in Fig. 2(c) for the free-running and locking regimes show that the second loop provides an efficient noise suppression up to about 30 kHz. Extending further the control bandwidth resulted in an unstable behavior due to the rather high 1.6 µs delay introduced by the AOFS, which is responsible for the servo-bump at 100 kHz and for the excess noise beyond it. Over a measurement time of 100 ms, the counted beat note suffers from an rms fluctuation of 100 and 35 kHz, respectively, for the slow and fast lock.

The higher frequency stability and spectral resolution afforded by the fast locking were found to be traded off by a severely degraded intensity noise. The oscilloscope traces reported in Fig. 2(d) show that the intensity drops by more than 50%. These oscillations emerge because the frequency jitter of the laser forces the AOFS to work outside its modulation bandwidth, i.e., at frequencies where its diffraction efficiency is degraded. This is better quantified in Fig. 2(b) by the comparison between the beat note spectrum (blue curve) and the diffraction response of the AOFS (dashed grey curve), which have comparable widths. It is worth noting that the intensity noise deriving from such an issue also impacts the quality of the feedback locking loop and could not be trivially solved by adoption of a faster AOFS.

To benefit from a constant power level (blue trace in Fig. 2 (d)), absolute spectroscopy measurements were performed with the EC-QCL slowly locked to the comb, under two different frequency scanning regimes. The first exploits the rather loose locking given by the piezo and forces the EC-QCL to jump from one comb mode to the next one, which implies unlocking and relocking at every spectral step. This is a robust procedure due to the inherently large capture range favored by the large laser linewidth. It produces an evenly spaced frequency axis with spectral points at every $f_{rep}$ (100 MHz in our case).

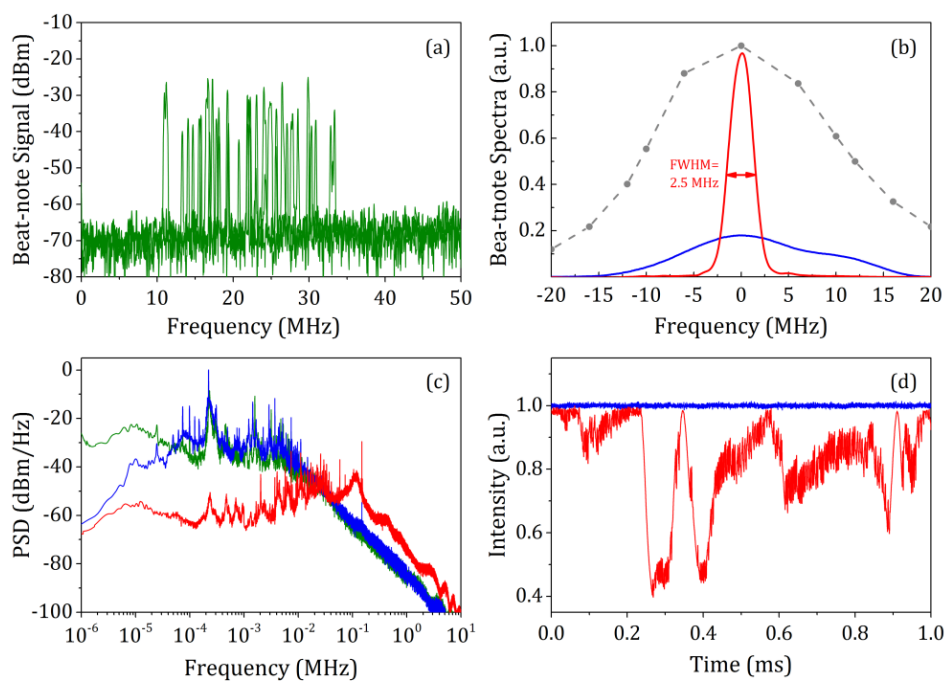

Fig. 2. Color code: green (free running), blue (slow locking), red (fast locking). (a) Beat-note signal spectrum acquired with a sweep time of 6 ms at a 50 kHz resolution bandwidth, showing a nearly 20 MHz large laser jittering window at a ms time scale. (b) Averaged electrical spectrum of the beat-note signal under slow and fast locking, as compared to the diffraction efficiency response of the AOFS (grey dashed dotted line). (c) Power spectral density of the error signal. (d) Scope traces of the laser intensity in locking condition

It can be applied for spectral scans up to 0.9 cm$^{-1}$, which is the limit given by the piezo. However, this approach can be easily extended to tens of wavenumbers with a remote control system that takes charge of driving both piezo and rotation stage of the laser.

To achieve a denser sampling of narrow spectral features, we tested both an interleaving of spectra acquired with different comb repetition rates and, as a second frequency scanning strategy, the tuning of the rep-rate while keeping a steady lock between EC-QCL and comb.

Figure 3(a) reports an example of absolute absorption spectrum near 1269 cm$^{-1}$ of an 85%-diluted $N_2O$ sample housed in a 66 cm long optical cell at a pressure of 0.25 mbar. The spectrum extends over 0.6 cm$^{-1}$ and presents 12.5 MHz-spaced points due to the interleaving of eight scans acquired at slightly detuned repetition frequencies (by ~30 Hz). The inset provides a zoomed-in view of the P(25) doublet and better highlights repeatability and absolute positioning of spectral points.

A quantitative analysis has been performed for the spectra reported in Fig. 3(b) and 3(c). The first refers to a 0.07%-diluted sample at 130 mbar. In this case, the 100 MHz sampling is sufficiently dense to reproduce the absorption spectrum and enables reliable fitting with a Voigt profile. On a statistical ensemble of 100 spectra, the fitting provides an rms deviation of 500 kHz for the line-center frequency, which is equivalent to 7 parts over $10^4$ with respect to a 750 MHz linewidth. The uncertainty primarily reflects the signal-to-noise-ratio (SNR) of the measurement, which amounts to ~1000 for the single spectrum. The residuals from the fitting do not show, at such a level of SNR, any appreciable departure from the Voigt profile. A more stringent test on precision (see Fig. 3 (c)) was obtained on the intense P(18) line of the 1000-0000 band at a pressure of 0.013 mbar, i.e., in conditions where the collisional broadening is negligible and the absorption linewidth is Doppler dominated to an estimated value of 70.5 MHz value. A 1.5 MHz spectral sampling is here ensured by a 4 Hz stepping of the comb rep-rate. This occurs at every 100 ms so that a 500 MHz large spectrum is acquired in 36 s. The statistical uncertainty on the line-center frequency, found by comparing consecutive back-and-forth spectral scans, is 70 kHz which is mainly limited in this case by the laser emission linewidth. In terms of systematic uncertainty, the limiting factor is related to an asymmetrical jittering of the laser around the local oscillator frequency, which translates in a beat note barycenter slightly detuned from the local oscillator itself (see Fig. 2(b)). Such detuning is accounted for by registering the electrical spectrum of the beat note signal during the spectral scan, but a residual systematic uncertainty at the 60 kHz level cannot be eliminated. This is prudent estimation that derives from the comparison of multiple spectra of the same line acquired in different conditions, changing the sign of the lock and also the local oscillator frequency. The resulting line center-frequency is 38052237297(62) kHz, the statistical uncertainty being almost negligible as compared to the systematic uncertainty due to sufficient averaging. The HITRAN value for the center-frequency is only 2.7 MHz above our determination and results more accurate than the nominal 3-30MHz confidence range. The Doppler width retrieved from the fitting is equal to 73.9 MHz, thus 2.4 MHz higher than the expected value, but this precisely reflects the instrumental broadening given by our 21 MHz large laser, once the typical quadrature addition law for Gaussian widths is applied.

In conclusion, we have reported on the architecture and performance of a novel comb-referred broadly-tunable laser source that provides 60 kHz accuracy levels in a spectral region where combs have failed so far to provide an impact on the spectroscopic knowledge encompassed in databases such as HITRAN. An effort is ongoing on the development of a fully-automated remote control system that is capable of fully exploiting the 1220-1325 cm$^{-1}$ tuning range of the EC-QCL. This will be applied to carry out accurate survey of P and R branches lines of the fundamental 1000-0000 band of $N_2O$. Other gas samples may interestingly be targeted in the currently available spectral region, such as $H_2O_2$ and $CH_4$, or in other regions till 12 μm by replacement of the laser-head.

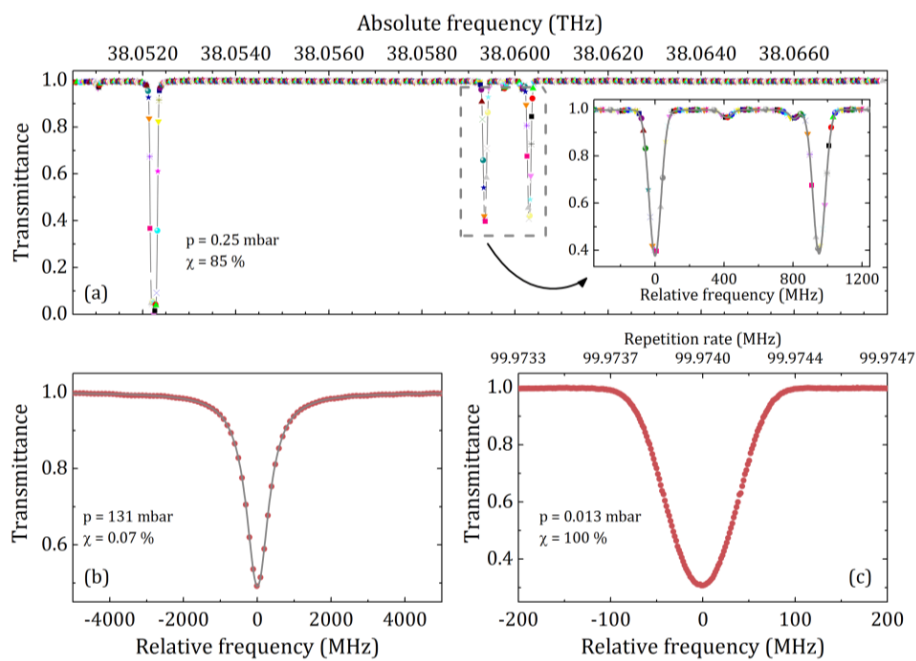

Fig. 3. (a): Absorption spectrum of an 85% nitrogen-diluted $N_2O$ sample at a pressure of 0.25 mbar near 1269 cm$^{-1}$, with a comb-defined frequency axis. Inset: zoomed-in view of the doublet, with interleaved spectra for a denser spectral sampling. (b) Absorption spectrum of the P(18) line of $N_2O$ with a 0.07% dilution at 131 mbar with a 100-MHz frequency grid. (c) Absorption spectrum of the same line in a pure sample at 0.013 mbar, here acquired by a 4 Hz stepping of the comb repetition frequency (1.5 MHz spaced optical frequency grid).

**Funding.** Funding received from the cooperative project FEAST between Politecnico di Milano and King Abdullah University of Science and Technology (KAUST) under the Competitive Center Funding program of KAUST and from the Italian Ministry of Research and Education (ELI project ESFRI Roadmap).

**REFERENCES**

1. R. Holzwarth, A.Yu. Nevsky, M. Zimmermann, Th. Udem, T.W. Hänsch, J. Von Zanthier, H. Walther, J.C. Knoght, W.J. Wadsworth, P.St.J. Russel, M.N. Skvortsov, S.N. Bagayev, Appl. Phys. B **73**, 269 (2001).
2. F.-L. Hong, A. Onae, J. Jiang, R. Guo, H. Inaba, K. Minoshima, T.R. Schibli, H. Matsumoto, Opt. Lett. **28**, 2324 (2003).
3. A. Amy-Klein, A. Goncharov, C. Daussy, C. Grain, O. Lopez, G. Santarelli, C. Chardonnet, Appl. Phys. B **78**, 25 (2004).
4. D. Mazzotti, P. Cancio, G. Giusfredi, De Natale, and M. Prevedelli, Opt. Lett. **30**, 997-999 (2005).
5. L. S. Rothman, C. P. Rinsland, A. Goldman, S. T. Massie, D. P. Edwards, J-M. Flaud, A. Perrin, C. Camy-Peyret, V. Dana, J.-Y. Mandin, J. Schroeder, A. Mccann, R. R. Gamache, R. B. Wattson, K. Yoshino, K. V. Chance, K. W. Jucks, L. R. Brown, V. Nemtchinov and P. Varanasi, J. Quant. Spectrosc. Radiat. Transfer **130**, 4 (2013).
6. C.S., Edwards, G. P. Barwood, H. S. Margolis, P. Gill, and W. R. C. Rowley, J Mol. Spectrosc. **234**, 143 (2005).
7. A. A. Madej, A. J. Alcock, A. Czajkowski, J. E. Bernard, and S. Chepurov, J. Opt. Soc. Am. B **23**, 2200 (2006).
8. S. Twagirayezu, M. J. Cich, T. J. Sears, C. P. McRaven, and G. E. Hall, J Mol. Spectrosc. **316**, 64 (2015).
9. A. Czajkowski, A. J. Alcock, J. E. Bernard, A. A. Madej, M. Corrigan, and S. Chepurov, Opt. Expr. **17**, 9258 (2009).
10. A. Gambetta, E. Fasci, A. Castrillo, M. Marangoni, G. Galzerano, G. Casa, P. Laporta and L. Gianfrani, New J. Phys. **12**, 103006 (2010).
11. D. Mondelain, T. Sala, S. Kassi, D. Romanini, M. Marangoni, and A. Campargue, J. Quant. Spectrosc. Radiat. Transfer **154**, 35 (2015).
12. S. M. Foreman, A. Marian, J. Ye, E. A. Petrukhin, M. A. Gubin, O. D. Mücke, F. N. C. Wong, E. P. Ippen, and F. X. Kärtner, Opt. Lett. **30**, 570 (2005).
13. A. Gambetta, M. Cassinerio, N. Coluccelli, E. Fasci, A. Castrillo, L. Gianfrani, D. Gatti, M. Marangoni, P. Laporta, and G. Galzerano, Opt. Lett. **40**, 304 (2015).
14. F. Adler, K. Cossel, M. J. Thorpe, I. Hart, M. E. Fermann and J. Ye, Opt. Lett. **34**, 1330-1332 (2009).
15. A. Amy-Klein, A. Goncharov, M. Guinet, C. Daussy, O. Lopez, A. Shelkovnikov, and C. Chardonnet, Opt. Lett. **30**, 3320 (2005).
16. S. Bartalini, P. Cancio, G. Giusfredi, D. Mazzotti, P. De Natale, S. Borri I. Galli, T. Leveque, and L. Gianfrani, Opt. Lett. **32**, 988 (2007).
17. D. Gatti, A. Gambetta, A. Castrillo, G. Galzerano, P. Laporta, L. Gianfrani, M. Marangoni, Opt. Expr. **19**, 17520 (2011).
18. J. Peltola, M. Vainio, T. Fordell, T. Hieta, M. Merimaa, and L. Halonen, Opt. Expr. **22**, 32429 (2014).
19. E. V. Kovalchuk, T. Schuldt, and A. Peters, Opt. Lett. **30**, 3141 (2005).
20. S Okubo, H. Nakayama, K. Iwakuni, H. Inaba, and H. Sasada, Opt. Expr. **19**, 23878 (2011).
21. I. Galli, S. Bartalini, P. C. Pastor, F. Cappelli, G. Giusfredi, D. Mazzotti, N. Akikusa, M. Yamanishi, and P. De Natale, Mol. Phys. **111**, 2041 (2013).
22. E. Baumann, F. R. Giorgetta, W. C. Swann, A. M. Zolot, I. Coddington, and N. R. Newbury, Phys Rev. A **84**, 062513 (2011).
23. K. Knabe, P. A. Williams, F. R. Giorgetta, C. M. Armacost, S. Crivello, M. B. Radunsky, and N. R. Newbury, Opt. Expr. **20**, 12432 (2012).s
24. K. Knabe, P. A. Williams, F. R. Giorgetta, M. B. Radunsky, C. M. Armacost, S. Crivello, and N. R. Newbury, Opt. Expr. **21**, 1020 (2013).
25. A. A. Mills, D. Gatti, J. Jiang, C. Mohr, W. Mefford, L. Gianfrani, M. Fermann, I. Hartl, and M. Marangoni, Opt. Lett. **37**, 4083 (2012)